\documentclass[12pt]{article} 
\usepackage[xdvi]{graphicx} 
\begin{document}   
\newcommand{\todo}[1]{{\em \small {#1}}\marginpar{$\Longleftarrow$}}   
\newcommand{\labell}[1]{\label{#1}\qquad_{#1}} %{\label{#1}} %  

\rightline{DCPT-03/41}   
\rightline{hep-th/0307216}   
\vskip 1cm

%% 			Title here  
%%  

\begin{center} 
{\Large \bf Properties of non-extremal enhan\c cons}
\end{center} 
\vskip 1cm   
  
\renewcommand{\thefootnote}{\fnsymbol{footnote}}   
\centerline{\bf   
Apostolos Dimitriadis\footnote{Apostolos.Dimitriadis@durham.ac.uk} and Simon 
F. Ross\footnote{S.F.Ross@durham.ac.uk}}    
\vskip .5cm   
\centerline{ \it Centre for Particle Theory, Department of  
Mathematical Sciences}   
\centerline{\it University of Durham, South Road, Durham DH1 3LE, U.K.}   
  
\setcounter{footnote}{0}   
\renewcommand{\thefootnote}{\arabic{footnote}}

%%			Text starts here  
%% 

\begin{abstract}   
We study the supergravity solutions describing non-extremal enhan\c
cons. There are two branches of solutions: a `shell branch' connected
to the extremal solution, and a `horizon branch' which connects to the
Schwarzschild black hole at large mass. We show that the shell branch
solutions violate the weak energy condition, and are hence
unphysical. We investigate linearized perturbations of the horizon
branch and the extremal solution numerically, completing an
investigation initiated in a previous paper. We show that these
solutions are stable against the perturbations we consider. This
provides further evidence that these latter supergravity solutions are
capturing some of the true physics of the enhan\c con.
\end{abstract}

\section{Introduction}     

The enhan\c con mechanism~\cite{JPP:enh} provides a novel
singularity-resolution mechanism in string theory. As such, it extends
the class of physical situations we can understand in terms of
weakly-coupled effective theories (this leads to interesting
applications in the AdS/CFT context~\cite{buchel:enh,evans:enh}), and
it allows us to explore how stringy physics changes our picture of the
structure of spacetime. The singularity resolution was originally
exhibited in~\cite{JPP:enh} for a BPS geometry, the repulson
singularity~\cite{Behrndt:enh,Kallosh:enh}, but both these
considerations motivate an interest in exploring the generalisation of
this mechanism to non-BPS, finite temperature solutions.

The question of the non-extremal generalisation of the enhan\c con
solution was briefly discussed in~\cite{JPP:enh}, and was considered
in more detail from a supergravity point of view in~\cite{JMPR:enh},
where it was found that there are two branches of non-extremal
solutions. One of them approaches the usual enhan\c con in the
extremal limit, and always has an enhan\c con shell, which may or may
not have a black hole living inside it; we refer to this as the shell
branch. The other branch has a smooth event horizon, and approaches
the Schwarzschild geometry far from extremality; we call this the
horizon branch. The shell branch exists for all masses greater than
the BPS value, but for the standard extremal solution, the horizon
branch exists only if the mass difference from the extremal solution
is greater than a certain critical value.

Physically, one expects that adding a small charge to a large black
hole should not drastically change the physics; therefore, it seems
that the horizon branch is the physically relevant solution for masses
much greater than the extremal value. However, there are no horizon
branch solutions for small enough mass difference. Thus, sufficiently
close to extremality, we should find a thermally-excited enhan\c con
shell. If this were to be described by some shell branch solution, we
would then expect that there should be transitions between the two
branches as we vary the parameters. This could provide a useful
example of the kind of non-trivial phase structure we expect to see in
such less supersymmetric contexts. Since we have the explicit
solutions, it should be possible to understand the phase structure in
detail. In a previous paper~\cite{es1}, we initiated an investigation
of the phase structure for these solutions, studying the
thermodynamics of the two branches and investigating the linearised
perturbations to see if there might be a classical instability which
could provide the mechanism for transitions between them.

In this paper, we will further investigate the physics of these
solutions. We discover that the story is quite different from what we
had expected. We will see that the shell branch of the non--extremal
enhan\c con violates the weak energy condition. We therefore regard
that branch as completely unphysical.\footnote{We would still expect
that one can slightly excite the enhan\c con shell; it is possible
that this is described by some more general ansatz within
supergravity. Alternatively, thermal excitation might smear the branes
out so that a thin shell is not a good approximation to the
distribution of branes, implying that no supergravity description is
possible in this regime. We will see that features of the shell branch
solutions argue in favor of the latter possibility.}

The shell branch does not represent real physics of the enhan\c con;
what about the horizon branch? To further investigate this, we have
completed our previous study of the linearised perturbations, finding
that there is no instability on the horizon branch to perturbations of
the type we consider. This provides some evidence that the horizon
branch is the physically correct description for the range of
parameters where it exists. We have also analysed the stability of the
extremal solution, determining appropriate boundary conditions at the
enhan\c con shell to supplement the linearised perturbation equations
obtained in~\cite{es1}. We do not find any instability here either; in
this case, this is very much the expected result. A BPS solution
should not have any instabilities. 

We should stress, however, that in neither case have we demonstrated
the absence of an instability; we have merely failed to find one
within the class of modes we studied. In particular, it is worth
noting that we have ignored the non-abelian fields coming from
D2-branes, which become light near the enhan\c con. These might play a
central role in the search for some more general solution in the low
mass difference region (a discussion in the extremal case has appeared
in~\cite{wijnholt}), and it might be interesting to extend the
perturbative analysis to include these fields, as a step towards
constructing such a solution. 

In section~\ref{sec:rev1}, we review the enhan\c con solutions,
extremal and non--extremal. We then show that the shell branch of the
non--extremal enhan\c con violates the weak energy condition in
section~\ref{sec:wec}.  In section~\ref{sec:rev2}, we review the
linearised perturbation analysis for the extremal and non-extremal
enhan\c con solutions carried out in~\cite{es1}. In
section~\ref{sec:hor}, we study the stability of the horizon
branch. We find no instability to the perturbations we are
considering. In section~\ref{sec:ext}, we determine the appropriate
matching conditions at the shell and carry out the relaxation of the
equations in the exterior region for the extreme case. This section
corrects flaws in the analysis of this case previously carried out
in~\cite{Maeda:stab}. Again, no instability is found.

\section{The enhan\c con solutions}
\label{sec:rev1}

The metrics we wish to study are the extremal and non--extremal
enhan\c con solutions obtained in~\cite{JPP:enh,JMPR:enh}.  These
geometries describe $N$ D$(p+4)$-branes wrapped on a K3; as
in~\cite{es1}, we focus on the case of two non-compact dimensions,
$p=2$. For the extremal case, the Einstein frame metric and fields are
\begin{eqnarray}
ds^2 &=& Z_2^{-5/8} Z_6^{-1/8} \eta_{\mu\nu} dx^\mu dx^\nu +
Z_2^{3/8} Z_6^{7/8}(dr^2 + r^2 d\Omega)  + V^{1/2} Z_2^{3/8}
Z_6^{-1/8} ds^2_{\rm K3} \ 
,\nonumber
\\
e^{2\phi } &=& g_s^2 {Z_2}^{1/2}{ Z_6}^{-3/2}\ , \nonumber\\
C_{(3)} &=& ({Z_2} g_s)^{-1} dx^0 \wedge dx^1 \wedge dx^2\ , \nonumber\\
C_{(7)} &=& ({Z_6} g_s)^{-1} dx^0 \wedge dx^1 \wedge dx^2 \wedge
V\,\varepsilon_{\rm K3}
\ , \label{outside}
\end{eqnarray}
where the harmonic functions are 
\begin{eqnarray}
Z_6 &=& 1+{r_6\over r}\ ,\quad Z_2 =1 +{r_2\over r}\ ,
\label{harms}
\end{eqnarray}
the parameters are related by\footnote{We are focusing on the simplest
case where there are no additional D2-branes.}
\begin{eqnarray}
\quad r_6 = {g_sN\alpha'^{1/2}\over 2} \ ,\quad r_2=-{V_*\over V} {r_6} \ ,
\label{arrs}
\end{eqnarray}
so $r_2$ is negative with our sign conventions, and $d\Omega$ denotes
the metric on the unit two-sphere.  The running K3 volume is
\begin{equation}
V(r)=V{Z_2(r)\over Z_6(r)}\ .
\end{equation}
$V = V_* = (2\pi \sqrt{\alpha'})^4$ at the enhan\c con radius, 
\begin{equation}
 r_{\rm e} = {2V_*\over {V} - V_* } r_6 \ .
\label{theradius}
\end{equation}
The repulson singularity would occur at $r= -r_2 < r_{\rm e}$. In the
enhan\c con mechanism~\cite{JPP:enh}, this singularity is resolved by
replacing the geometry inside the enhan\c con radius identified above
by a flat space. The geometry inside the shell is
\begin{equation}
g_s^{1/2}\,ds^2 = H_2^{-5/8} H_6^{-1/8} \eta_{\mu\nu} dx^\mu dx^\nu +
H_2^{3/8} H_6^{7/8} (dr^2 + r^2 d\Omega) + V^{1/2}
H_2^{3/8} H_6^{-1/8} ds^2_{K3}
\end{equation}
and the non--trivial fields are
\begin{eqnarray}
e^{2\phi} &=& g_s^2 H_2^{1/2} H_6^{-3/2}\ , \nonumber\\
C_{(3)} &=& (g_s H_2)^{-1} dx^0
\wedge dx^1 \wedge dx^2\ ,\nonumber\\ 
C_{(7)} &=& (g_s H_6)^{-1} dx^0\wedge dx^1 \wedge dx^2
\wedge V\,\varepsilon_{\rm K3}\ ,
\end{eqnarray}
where
\begin{eqnarray}
H_2 = 1+{r_2 \over r_{\rm e}},
 \label{h2}\\
H_6 = 1+{r_6\over r_{\rm e}} .
 \label{h6}
\end{eqnarray}
These constants are introduced to make the continuity of the solution
at the shell explicit. 

In~\cite{JMPR:enh}, it was found that this extremal solution could be
generalised to obtain two branches of non--extremal solutions, arising
from an ambiguity of a choice of sign in the solution of the
supergravity equations for the usual ansatz. The non-extremal
generalisation of the exterior geometry is
\begin{eqnarray}
g_s^{1/2}\,ds^2  &=&Z_2^{-5/8}{Z}_6^{-1/8} (-K dt^2 + dx_1^2+ dx_2^2) +
Z_2^{3/8}Z_6^{7/8} (K^{-1} dr^2 + r^2 d\Omega_2^2) \nonumber\\
&\phantom{+}&\hskip5cm + V^{1/2} Z_2^{3/8}Z_6^{-1/8}ds_{\rm K3}^2\ ,
\label{junterior}
\end{eqnarray}
the dilaton and R--R fields are 
\begin{eqnarray}
e^{2\phi} &=& g_s^2Z_2^{1/2}Z_6^{-3/2}\ , \nonumber \\
C_{(3)} &=& (g_s \alpha_2 Z_2)^{-1}dt \wedge dx^1 \wedge dx^2\ ,
\nonumber\\
C_{(7)} &=& (g_s \alpha_6 Z_6)^{-1} dt \wedge dx^1 \wedge dx^2 \wedge
V\,\varepsilon_{\rm K3}\ , \label{fields}
\end{eqnarray}
and the various harmonic functions are given by
\begin{eqnarray}
K &=& 1 - {r_0 \over r}\ ,\nonumber\\ %\labell{nonxtre}\\
Z_2&=&1 + {\hat{r}_2 \over r}\ \qquad Z_6= 1 + {\hat{r}_6 \over r}\ .
\label{z6ext}
\end{eqnarray}
Here 
\begin{equation}
\hat{r}_6 = - {r_0 \over 2} + \sqrt{r_6^2 + \left( {r_0 \over 2} \right)^2}\ ,
\end{equation}
and $\alpha_6 = \hat{r}_6/r_6$. There are two choices for $\hat{r}_2$
consistent with the equations of motion:
\begin{equation}
\hat{r}_2 =  - {r_0 \over 2} \pm
\sqrt{r_2^2 + \left( {r_0 \over 2} \right)^2}\ ,
\label{choice}
\end{equation}
and $\alpha_2 = \hat{r}_2/r_2$. Here, $r_2$ and $r_6$ are still given
by (\ref{arrs}).

There are two branches of solutions. For the upper sign in
(\ref{choice}), $\hat{r}_2 >0$, so there is no repulson singularity,
and the solution has a regular horizon at $r=r_0$. For the lower
choice of sign, however, the repulson singularity always lies outside
the would-be horizon, $|\hat{r}_2| > r_0$, and the geometry will be
corrected by an enhan\c con shell. We therefore refer to the former as
the horizon branch and the latter as the shell branch. 

The shell branch exterior solution is cut off by an enhan\c con shell at
\begin{equation}
r_{\rm e} = {V_* \hat{r}_6 - V \hat{r}_2 \over V - V_*}\ .
\label{junktre}
\end{equation}
The interior solution can in general contain an uncharged black hole, 
\begin{eqnarray}
g_s^{1/2}\,ds^2 &=& H_2^{-5/8} H_6^{-1/8} \left(- {K(r_{\rm e})\over
L(r_{\rm e})}L dt^2 + dx_1^2+dx_2^2\right) + H_2^{3/8} H_6^{7/8}
(L^{-1} dr^2 + r^2 d\Omega) \nonumber\\
&\phantom{+}&\hskip5cm + V^{1/2} H_2^{3/8} H_6^{-1/8}ds^2_{K3}\ ,
\label{tag1}
\end{eqnarray}
with accompanying fields
\begin{eqnarray}
e^{2\phi} &=& g_s^2 H_2^{1/2} H_6^{-3/2}\ , \nonumber\\
 \qquad C_{(3)} &=&
\left({K(r_{\rm e})\over L(r_{\rm e})}\right)^{1/2} (g_s
 \alpha'_2H_2)^{-1} dt \wedge dx^1 \wedge dx^2\ , \nonumber\\
C_{(7)} &=& \left({K(r_{\rm e})\over L(r_{\rm e})}\right)^{1/2}
(g_s \alpha'_6 H_6)^{-1} dt\wedge dx^1 \wedge dx^2 
\wedge V\,\varepsilon_{\rm K3}\ ,
\label{infields}
\end{eqnarray}
where
\begin{eqnarray}
L &=& 1 - {r_0' \over r}\ ,\nonumber\\
H_2 &=& 1 +{\hat{r}_2 \over r_{\rm e}} ,
\nonumber\\
H_6 &=& 1 +{\hat{r}_6 \over r_{\rm e}} .
\label{tag2}
\end{eqnarray}
In this supergravity analysis, there is an independent
non--extremality scale $r_0'$ for the interior solution. It was argued
in~\cite{JMPR:enh} that this is an unphysical degree of freedom, which
would be determined uniquely (or possibly up to some discrete
ambiguity) in terms of the asymptotic mass and charges if we properly
understood the physics of the shell. Unfortunately, this purely
supergravity analysis provides too crude a description of the shell in
the non-extremal context to do this. All we can say is that
$r_0'<r_{\rm e}$, in order that the interior black hole actually fits
inside the shell.

\section{Shell branch violates weak energy condition\protect\footnote{This
    observation arose from collaborative discussions with Amanda Peet.} }
\label{sec:wec}

In fact it turns out that it is not just this freedom to specify
$r_0'$ which is unphysical: the shell branch solution given above is
unphysical for any value of $r_0'$, as the stress tensor of the shell
required to match the exterior and interior solutions violates the
weak energy condition. 

This is easily seen by considering the $tt$ component of the stress
tensor of the thin shell, resulting from the Israel junction
conditions~\cite{JMPR:enh}:
\begin{eqnarray}
2\kappa
^2S_{tt}=\frac{1}{\sqrt{G_{rr}}}\left[\frac{\hat{Z}'_2}{\hat{Z}_2}+\frac{\hat{Z}'_6}{\hat{Z}_6}
+\frac{4}{r_{\rm i}}\left(1-\sqrt{\frac{L(r)}{K(r)}}\right)
\right]G_{tt}
\label{nonextStt}
\end{eqnarray}
This determines the energy density $\rho$ of the thin
shell, which scales as 
\begin{eqnarray}
\rho \sim -\frac{\hat{Z}'_2}{\hat{Z}_2}-\frac{\hat{Z}'_6}{\hat{Z}_6}
+ \frac{4}{r_{\rm i}}\left(\sqrt{\frac{L(r)}{K(r)}}-1\right)
\label{nonextrho}
\end{eqnarray}
The second term is always positive, as $\hat r_6 >0$, implying that
$\hat Z_6' <0$ (that is, $\hat Z_6$ is decreasing as $r$
increases). We generally assume that $r_0' < r_0$, so $L(r) > K(r) $,
implying that the third term is positive. However, for solutions on
the shell branch, $\hat r_2 <0$, so the first term is negative.  

If we consider a shell at large radius, 
\begin{equation}
\rho \sim {\hat r_2 + \hat r_6 + 2(r_0 - r_0') \over r^2},
\end{equation}
and the negative contribution from $\hat r_2$ is always balanced by
the positive contributions from the other terms, to give a positive
answer. This is just another way of saying that the exterior solution
has a positive ADM mass (in fact, its ADM mass is greater than the
mass of the extremal solution). 

However, this is not what we want to consider. The radius of the shell
is supposed to be close to the enhan\c con radius. The physical
argument for this is that the solution is constructed by bringing
branes in from infinity. In the extremal case, the branes see no
potential, but the closest radius we can bring them to is the enhan\c
con radius, because they cease to behave like pointlike branes at this
radius \cite{JPP:enh}. In the non-extremal case, the branes feel an
attractive force in the exterior geometry \cite{JMPR:enh}, so they
will again not stop until they start to smear out at the enhan\c con
radius.

When we consider a shell at the enhan\c con radius, the energy density
$\rho$ will be negative. At the enhan\c con radius, ${\hat{Z}_2 \over
\hat{Z}_6}= {V_* \over V}$, so $\rho$ in (\ref{nonextrho}) can be
rewritten as
\begin{eqnarray}
\rho \sim \frac{\hat{r}_2}{r_{\rm
e}}\frac{1}{\hat{Z}_2}\left(1+\frac{V_*}{V}\frac{\hat{r}_6}{\hat{r}_2}\right)
+\frac{4}{r_{\rm e}}\left(\sqrt{\frac{L(r_{\rm e})}{K(r_{\rm
e})}}-1\right).
\label{rho}
\end{eqnarray}
On the shell branch, $\hat r_2$ is negative, and $|\hat r_6/ \hat r_2|
< V/V_*$, so the first term is negative. When $V_* / V$ is small,
$\hat Z_2$ is small, and the first term will dominate over the second,
so that the overall energy density of the shell will be negative. Now
we need $V \gg V_*$ for this supergravity analysis to be relevant; so
in the regime where this description is supposed to apply, the energy
density of the shell is negative.

One can extend this discussion to consider solutions on the shell
branch with additional D2-brane charge, as considered
in~\cite{JMPR:enh}. The expression for the energy density of the shell
becomes more complicated in this case, and we have not been able to
find a simple argument that it will always be negative, but numerical
investigation shows that the energy density of the shell is negative
for all values the parameters we tried also in this more general case.

Thus, the weak energy condition is violated on the shell branch. This
implies that the shell branch solution does not correctly describe
small perturbations away from the enhan\c con solution; one
possibility is that this signals a breakdown of the thin-shell
approximation for small departures from extremality. There is some
evidence for this interpretation coming from the study of probe
branes. In the BPS case considered previously, we had a moduli space
of solutions of the classical equations of motions, and we argued that
we could choose all the sources to lie at as small a radius as
possible, justifying the thin shell picture. In thermal equilibrium at
some non-zero temperature, the constituent branes will carry thermal
kinetic energy, and it is not clear that the inter-brane interactions
will be sufficient to restrain the branes within a narrow range in
$r$. Near extremality, the average extra energy per brane scales as
$r_0/N$, while the typical scale of the effective potential seen by a
probe brane in the exterior region is $V_{eff} \sim r_0/r_{\rm e} \sim
r_0/N$, so it is not clear that branes in the enhan\c con shell will
remain confined to a thin layer when we add some thermal
energy.\footnote{We are grateful to Rob Myers for discussions on this
point.}

\section{Linearised perturbation equations}
\label{sec:rev2}

Having shown that the shell branch solutions are unphysical, we would
now like to turn our attention to what more we can learn about the
other solutions. To further explore the physics of the horizon branch
and the extremal metric, it is interesting to continue the analysis of
small perturbations of these solutions initiated in~\cite{es1}.

In~\cite{es1}, we considered the linearised perturbations of the
exterior geometry, assuming that the perturbation preserves the
invariance under spherical symmetry in the $\theta, \phi$ directions,
translational invariance in $x_1$ and $x_2$, and the discrete
symmetries under $x_1 \to -x_1$, $x_2 \to -x_2$, $\phi \to -\phi$. By
a suitable choice of coordinates, the most general perturbation
consistent with these symmetries can be written as the metric
\begin{eqnarray}
g_s^{1/2}\,ds^2  &=& e^{-\psi_1/2} \left[ \bar{Z}_2^{-1/2}
\bar{Z}_6^{-1/2} (-\bar{K} 
e^{\delta \psi_2} dt^2 + e^{-{1 \over 2} \delta \psi_2+ \delta \psi_3}
dx_1^2 + e^{-{1\over 2} \delta \psi_2 - \delta \psi_3} dx_2^2)
\right. \nonumber \\ &\phantom{+}& \left. +\bar{Z}_2^{1/2}
\bar{Z}_6^{1/2} (\bar{K}^{-1} dr^2 + r^2 d\Omega_2^2)  + V^{1/2}
\bar{Z}_2^{1/2} \bar{Z}_6^{-1/2}ds_{\rm K3}^2\right] \ ,
\label{pertext}
\end{eqnarray}
dilaton 
\begin{equation}
\bar{\phi} = \phi + \delta \phi, 
\end{equation}
and R--R fields 
\begin{equation}
\bar{C}_{(3)} = C_{(3)} + \delta C_{(3)}, \quad \bar{C}_{(7)} = C_{(7)} +
\delta{C}_{(7)}.   
\end{equation}
Here 
\begin{equation}
\psi_1 = \phi + \delta \psi_1,\  \bar{Z}_2 = Z_2 (1 + \delta Z_2),\ 
\bar{Z}_6 = Z_6 (1 + \delta Z_6),\  \bar{K} = K (1 + \delta K),
\end{equation}
the harmonic functions $Z_2, Z_6, K$ are as in (\ref{z6ext}), the
unperturbed dilaton $\phi$ is as in (\ref{fields}), and the RR
potentials are as in (\ref{fields}). The first-order perturbations are
all functions of $(r,t)$ only, while the background quantities are
functions only of $r$.  We look for perturbations of the form $f(r)
e^{\sigma t}$.\footnote{Note that our ansatz is slightly more general
than the ansatz adopted in the study of perturbations of the extremal
enhan\c con geometry in~\cite{Maeda:stab}. We introduced three new
perturbation functions, $\delta \psi_2$, $\delta \psi_3$, and $\delta
K$. We can choose to set $\delta K=0$ by a gauge transformation, and
$\delta \psi_3$ decouples---its analysis was completed in~\cite{es1},
so we will set $\delta \psi_3=0$ in this paper. However, the
additional function $\delta \psi_2$ is needed to solve the full set of
linearised field equations, even in the extremal case.}

Writing the metric in the form (\ref{pertext}) does not fix the
diffeomorphism freedom completely.  There are
infinitesimal diffeomorphisms which preserve its form. Namely,
\begin{equation}
t \to t' = t + e^{\sigma t} \delta t(r), \quad r \to r' = r +
e^{\sigma t} \delta r(r),
\end{equation}
with 
\begin{equation}  
\partial_r \delta t = \sigma {Z_2 Z_6 \over K^2} \delta r . 
\end{equation}
This diffeomorphism contains an arbitrary function; since we are not
interested in pure gauge perturbations, we should fix this additional
gauge symmetry. We can do so by setting one of the perturbations to
zero. It is convenient to choose $\delta K=0$. There remain
diffeomorphisms which will preserve $\delta K=0$: these have
\begin{equation} \label{rdif1}
\delta r = a r K^{1/2}
\end{equation}
and
\begin{eqnarray} \label{rdif2}
\delta t &=& \sigma a
\left[-\frac{2(r_0+\hat{r}_2)(r_0+\hat{r}_6)}{\sqrt{K(r)}}+\left(\frac{r}{2}+\frac{7r_0}{4}+\hat{r}_2
+\hat{r}_6\right)r\sqrt{K(r)}+\right. \\ \nonumber
&&\left.+\left(\frac{15r_0^2}{8}+\frac{3r_0(\hat{r}_2+\hat{r}_6)}{2}+\hat{r}_2\hat{r}_6\right)
\ln{\frac{1+\sqrt{K(r)}}{1-\sqrt{K(r)}}} \right]+ \sigma b
\end{eqnarray}
If we apply this diffeomorphism to the non-extremal enhan\c con geometry
(\ref{junterior}), we obtain a metric of the form (\ref{pertext}) with
\begin{eqnarray} \label{partint}
\delta \psi_1^d &=& \left( \phi' - {4 \over 3r} \right) \delta r - {2
\over 3} \partial_r \delta r - {2 \over 3} \sigma \delta t, \\
\delta \psi_2^d &=& - {4 \over 3r} \delta r + {4 \over 3} \partial_r
\delta r + {4 \over 3} \sigma \delta t, \nonumber \\
\delta Z_6^d &=& \left( {Z_6' \over Z_6} + {2 \over r} \right) \delta r,
\nonumber \\
\delta Z_2^d &=& \left( {Z_2' \over Z_2} +{2 \over 3r} \right) \delta r
- {2 \over 3} \partial_r \delta r - {2 \over 3} \sigma \delta t,
\nonumber \\
\delta \phi^d &=& \phi' \delta r .\nonumber
\end{eqnarray}

In~\cite{es1}, we showed that if we replace the constants $a$ and $b$
by functions $a(r)$ and $b(r)$, and rewrite the general perturbation
as 
\begin{eqnarray}
\delta \psi_1 &=& \delta \psi_1^d(a(r),b(r)), \\
\delta \psi_2 &=& \delta \psi_2^d(a(r),b(r)) + \Psi_2, \nonumber \\
\delta Z_6 &=& \delta Z_6^d(a(r),b(r)) + {\cal Z}_6, \nonumber \\
\delta Z_2 &=& \delta Z_2^d(a(r),b(r)), \nonumber \\
\delta \phi &=& \delta \phi^d(a(r),b(r)) + \Phi \nonumber,
\end{eqnarray}
the full set of linearised equations of motion reduce to two algebraic
equations for $\partial_r a$ and $\partial_r b$ and the following
system of second-order equations for the independent functions
$\Psi_2, {\cal Z}_6, \Phi$ (where $'$ again denotes $\partial_r$, and we
assume that all the perturbations behave as $e^{\sigma t}$,
representing an instability). There
is an equation involving $\Phi''$,
\begin{equation} \label{peq1}
D (\Phi'' + {2r-r_0 \over r^2 K} \Phi' - {Z_2 Z_6 \over K^2} \sigma^2
\Phi) + P^1_2 (\Psi_2' + 2 {\cal Z}_6') + Q^1_1 \Phi + Q^1_2 \Psi_2 +
Q^1_3 {\cal Z}_6 =0,
\end{equation}
with the polynomial coefficients
\begin{equation}
D = r^2 K (8 r^2 + 5
r \hat{r}_2  + 5 r \hat{r}_6 + 2 \hat{r}_2 \hat{r}_6 ) (4 r^2 + 3 r
\hat{r}_2 + 3 r\hat{r}_6  + 2 \hat{r}_2 \hat{r}_6),
\end{equation}
\begin{equation}
P^1_2 = -2 r^2 K ( - 2 r^2 \hat{r}_2 + 6r^2 \hat{r}_6 + 8 r \hat{r}_2 \hat{r}_6
+ 3 \hat{r}_2^2 \hat{r}_6  +  \hat{r}_2 \hat{r}_6^2 ),
\end{equation}
\begin{eqnarray}
Q^1_1 &=& -r^2 (4 r_0 \hat{r}_2 + 36 r_0 \hat{r}_6 + 3 \hat{r}_2^2 + 6
\hat{r}_2 \hat{r}_6 + 27 \hat{r}_6^2) \\ && \nonumber - r (40 r_0
\hat{r}_2 \hat{r}_6 
+ 2 \hat{r}_2^2 \hat{r}_6 +  30 \hat{r}_2 \hat{r}_6^2) 
-  12 r_0 \hat{r}_2^2 \hat{r}_6 - 8 \hat{r}_2^2 \hat{r}_6^2,
\end{eqnarray}
\begin{equation}
Q^1_2 = r_0 (-2 r^2 \hat{r}_2 + 6 r^2 \hat{r}_6 + 8 r \hat{r}_2
\hat{r}_6 + 3 \hat{r}_2^2 \hat{r}_6 + \hat{r}_2 \hat{r}_6^2), 
\end{equation}
\begin{eqnarray}
Q^1_3 &=& r^2 (8 r_0 \hat{r}_2 + 24 r_0 \hat{r}_6 + 9 \hat{r}_2^2 + 10
\hat{r}_2 \hat{r}_6 + 9 \hat{r}_6^2) \\ && \nonumber + r (24 r_0
\hat{r}_2 \hat{r}_6 + 
6 \hat{r}_2^2 \hat{r}_6 + 10 \hat{r}_2 \hat{r}_6^2)  + 6 r_0
\hat{r}_2^2 \hat{r}_6 -2 r_0 \hat{r}_2 \hat{r}_6^2.
\end{eqnarray}
The equation involving $\Psi_2''$ is
\begin{equation} \label{peq2}
D (\Psi_2'' - {Z_2 Z_6 \over K^2} \sigma^2 \Psi_2) + P^2_2 \Psi_2' + P^2_3
{\cal Z}_6' + Q^2_1 \Phi + Q^2_2 \Psi_2 + Q^2_3 {\cal Z}_6 = 0,
\end{equation}
where $D$ is as before, and the other polynomial  coefficients are
\begin{eqnarray}
P^2_2 &=& 64 r^5 + r^4(-32 r_0 + 120 \hat{r}_2 + 88\hat{r}_6) \\ &&
\nonumber + r^3 (-76
r_0  \hat{r}_2 -44 r_0 \hat{r}_6 +30 \hat{r}_2^2 + 172 \hat{r}_2
\hat{r}_6 + 30 \hat{r}_6^2) \\ && \nonumber + r^2 (-15 r_0 \hat{r}_2^2 -118 r_0
\hat{r}_2 \hat{r}_6 -15 r_0 \hat{r}_6^2 + 44 \hat{r}_2^2 \hat{r}_6 +
52 \hat{r}_2 \hat{r}_6^2) \\ && \nonumber + r (-28 r_0 \hat{r}_2^2 \hat{r}_6 -36 r_0
\hat{r}_2 \hat{r}_6^2 + 8 \hat{r}_2^2 \hat{r}_6^2) - 4 r_0 \hat{r}_2^2
\hat{r}_6^2 ,
\end{eqnarray}
\begin{equation}
P^2_3 = -8 r^2 \hat{r}_2 K (8r^2 + 16 r \hat{r}_6 + 3 \hat{r}_2 \hat{r}_6 +
5 \hat{r}_6^2), 
\end{equation}
\begin{equation}
Q^2_1 = 4 \hat{r}_2 (r^2 (-8 r_0 - 6 \hat{r}_2 -6 \hat{r}_6) + r (4
r_0 \hat{r}_6 - 7 \hat{r}_2 \hat{r}_6 + 3 \hat{r}_6^2) + 6 r_0
\hat{r}_2 \hat{r}_6 +2 \hat{r}_2\hat{r}_6^2),
\end{equation}
\begin{equation}
Q^2_2 = -2 r_0 \hat{r}_2 (8r^2 + 16 r \hat{r}_6 + 3 \hat{r}_2
\hat{r}_6 + 5 \hat{r}_6^2),
\end{equation}
\begin{equation}
Q^2_3 = 4 \hat{r}_2 (r^2(16 r_0 + 18 \hat{r}_2 + 2 \hat{r}_6) + r (12
r_0 \hat{r}_6  +21 \hat{r}_2 \hat{r}_6 - \hat{r}_6^2) -3 r_0 \hat{r}_2
\hat{r}_6 + 5 r_0 \hat{r}_6^2 + 6 \hat{r}_2 \hat{r}_6^2).
\end{equation}
The equation involving ${\cal Z}_6''$ is 
\begin{equation} \label{peq3}
D ({\cal Z}_6'' - {Z_2 Z_6 \over K^2} \sigma^2 {\cal Z}_6) + P^3_2
\Psi_2' + P^3_3 {\cal Z}_6' + Q^3_1 \Phi + Q^3_2 \Psi_2 + Q^3_3 {\cal
Z}_6 =0,
\end{equation}
where $D$ is as before, and the other polynomial  coefficients are
\begin{equation}
P^3_2 = -2r^2 K (6 r^2 \hat{r}_2 -2 r^2 \hat{r}_6 + 8 r \hat{r}_2
\hat{r}_6 + \hat{r}_2^2 \hat{r}_6 + 3 \hat{r}_2\hat{r}_6^2),
\end{equation}
\begin{eqnarray}
P^3_3 &=& 64 r^5 + r^4 (-32 r_0 + 64 \hat{r}_2 + 96 \hat{r}_6) \\ &&
\nonumber + r^3
(-20 r_0 \hat{r}_2 -52 r_0 \hat{r}_6 + 30 \hat{r}_2^2 +76 \hat{r}_2
\hat{r}_6 + 30 \hat{r}_6^2) \\ \nonumber &&+ r^2 (-15 r_0 \hat{r}_2^2 -22 r_0
\hat{r}_2 \hat{r}_6 -15 r_0 \hat{r}_6^2 + 28 \hat{r}_2^2 \hat{r}_6 +
20 \hat{r}_2 \hat{r}_6^2) \\ \nonumber && + r (-12 r_0 \hat{r}_2^2
\hat{r}_6 -4 r_0 
\hat{r}_2 \hat{r}_6^2 + 8 \hat{r}_2^2 \hat{r}_6^2) -4 r_0 \hat{r}_2^2
\hat{r}_6^2,
\end{eqnarray}
\begin{equation}
Q^3_1 = r^2(12 r_0 \hat{r}_2 + 12 r_0 \hat{r}_6 + 9 \hat{r}_2^2 + 10
\hat{r}_2 \hat{r}_6 + 9 \hat{r}_6^2) + r ( 8 r_0 \hat{r}_2 \hat{r}_6 +
10 \hat{r}_2^2 \hat{r}_6 + 6\hat{r}_2 \hat{r}_6^2) -4 r_0 \hat{r}_2^2
\hat{r}_6, 
\end{equation}
\begin{equation}
Q^3_2 = r_0 (6 r^2 \hat{r}_2 -2 r^2 \hat{r}_6 + 8 r \hat{r}_2
\hat{r}_6 + \hat{r}_2^2 \hat{r}_6 + 3 \hat{r}_2 \hat{r}_6^2),
\end{equation}
\begin{eqnarray}
Q^3_3 &=& -r^2 (24 r_0 \hat{r}_2 + 8 r_0 \hat{r}_6 + 27 \hat{r}_2^2 +
6 \hat{r}_2 \hat{r}_6 + 3 \hat{r}_6^2) \\ && \nonumber - r (24 r_0 \hat{r}_2 \hat{r}_6
+ 30 \hat{r}_2^2 \hat{r}_6 +2 \hat{r}_2 \hat{r}_6^2) + 2 r_0
\hat{r}_2^2 \hat{r}_6 - 6 r_0 \hat{r}_2 \hat{r}_6^2 - 8 \hat{r}_2^2
\hat{r}_6^2).  
\end{eqnarray}

The problem of finding unstable perturbations of the enhan\c con
geometries in our ansatz then reduces to finding solutions of this
coupled system of equations satisfying some appropriate boundary
conditions.

\section{Horizon-branch stability}
\label{sec:hor}

Let us first consider the perturbations for the horizon branch
solutions. The appropriate boundary conditions are then just that the
linearised perturbations should be regular on the horizon $r=r_0$ and
at infinity. The solutions of the equations
(\ref{peq1},\ref{peq2},\ref{peq3}) behave as
\begin{equation}
\Phi, \Psi_2, {\cal Z}_6 \to (r-r_0)^{\pm \bar \sigma}, \bar \sigma =
(r_0 + \hat r_2)^{1/2} (r_0 + \hat r_6)^{1/2} \sigma 
\end{equation}
as $r \to r_0$, and they behave as 
\begin{equation}
\Phi, \Psi_2, {\cal Z}_6 \to e^{\pm \sigma r}
\end{equation}
as $r \to \infty$. We wish to know if there is some $\sigma$ such that
we obtain a solution where $\Phi, \Psi_2, {\cal Z}_6$ all have decaying
behaviour both at infinity and the horizon.

We have investigated this question numerically, using a simple
relaxation method \cite{PPT:num} to search for solutions satisfying the boundary
conditions. We start from a trial solution satisfying the falloff
conditions at the horizon and infinity, with a smooth interpolation
with no nodes (as we are most interested in the instability with
largest $\sigma$, which we would expect to have no nodes). We then
relax $\Phi, \Psi_2, {\cal Z}_6, \sigma$ to see if we can reach a
solution of the equations of motion. We have explored a wide range of
the free parameters $\hat{r}_6, \hat{r}_2, r_0$ of the background
solution, and we never find any instability. The relaxation process
fails to converge. 

This is the expected result for large non-extremality; in this limit,
the horizon branch solutions approach a four-dimensional Schwarzschild
metric smeared over the K3 and longitudinal $x_1,x_2$ directions, and
this Schwarzschild metric is known to be stable against the kind of
perturbations we are considering \cite{chandra} (note that our
perturbations are assumed independent of the longitudinal coordinates,
so the Gregory-Laflamme instability\cite{GL} which will appear if the
$x_1,x_2$ are non-compact is absent from this analysis).

The non-trivial result is that this stability persists over the whole
of the horizon branch. Thus, the linearised stability analysis has
revealed no sign of any transition from this branch of solutions to
any other solution as the mass decreases. This is very interesting;
although it does not rule out such a transition, the horizon branch
has passed the first test we could subject it to, and provides the
best available description of the non-extremal enhan\c con physics in
the region where it is available. This leads us to suspect that there
is no other supergravity solution describing enhan\c con physics for
the range of parameters where the horizon branch solution exists.

\section{Extremal stability}
\label{sec:ext}

To search for instabilities in the extreme case, where there is an
with enhan\c con shell, we need to determine the appropriate matching
conditions at the shell for the linearised perturbations $\Phi,
\Psi_2, {\cal Z}_6$. Since these arise as perturbations of components
of the metric which have non-trivial matching conditions relating the
discontinuity of their derivatives to the shell stress-energy,
obtaining the appropriate matching conditions is a non-trivial
problem. We will use the DBI action to obtain these matching
conditions. The ensuing numerical study finds no instabilities for the
extremal case, as expected from supersymmetry arguments.

Outside the shell, we assume we have the perturbed metric
(\ref{pertext}), and the associated dilaton and RR fields, with
$r_0=0$, so $K=1$ (since we have used the diffeomorphism freedom to
set $\delta K=0$). Inside the shell, we will have a perturbed flat
space,
\begin{eqnarray}
g_s^{1/2}\,ds^2 &=& e^{-\gamma_1/2} \left[ \bar{H}_2^{-1/2}
\bar{H}_6^{-1/2} (- e^{\delta \gamma_2} dt^2 + e^{-{1 \over 2} \delta
\gamma_2} (dx_1^2 + dx_2^2)) \right. \nonumber \\ &\phantom{+}&
\left. +\bar{H}_2^{1/2} \bar{H}_6^{1/2} (dr^2 + r^2
d\Omega_2^2) + V^{1/2} \bar{H}_2^{1/2} \bar{H}_6^{-1/2}ds_{\rm
K3}^2\right] \ ,
\label{pertint}
\end{eqnarray}
dilaton 
\begin{equation}
\bar{\phi} = \phi + \delta \xi, 
\end{equation}
and R--R fields 
\begin{equation}
\bar{C}_{(3)} = C_{(3)} + \delta C_{(3)}, \quad \bar{C}_{(7)} = C_{(7)} +
\delta{C}_{(7)}.   
\end{equation}
Here 
\begin{equation}
\gamma_1 = \phi + \delta \gamma_1,\  \bar{H}_2 = H_2 (1 + \delta H_2),\ 
\bar{H}_6 = H_6 (1 + \delta H_6),
\end{equation}
the constants $H_2, H_6$ are as in (\ref{z6ext}) with $r_0'=0$, the
unperturbed dilaton $\phi$ is as in (\ref{infields}), and the RR
potentials are as in (\ref{infields}).

We can analyse the perturbations of the interior geometry following
the same route used above for the exterior geometry. The
diffeomorphisms preserving the form of our ansatz are now $\delta r =
c r$, $\delta t = \sigma H_2 H_6 c r^2/2 + \sigma d$. We can write 
\begin{eqnarray}
\delta \gamma_1 &=& \delta \gamma_1^d(c(r),d(r)), \\
\delta \gamma_2 &=& \delta \gamma_2^d(c(r),d(r)) + \Gamma_2, \nonumber \\
\delta H_6 &=& \delta H_6^d(c(r),d(r)) + {\cal H}_6, \nonumber \\
\delta H_2 &=& \delta H_2^d(c(r),d(r)), \nonumber \\
\delta \xi &=& \delta \xi^d(c(r),d(r)) + \Xi \nonumber,
\end{eqnarray}
and then we find that the equations for the free perturbation
functions are all
\begin{equation}
\partial_r^2 f + {2 \over r} \partial_r f - H_2 H_6 \sigma^2 f = 0,
\end{equation}
where $f = \Xi, \Gamma_2, {\cal H}_6$. The solution regular at $r=0$ is
\begin{equation}
f = f_0 {\sinh \bar \sigma r \over r}, 
\end{equation}
where $\bar \sigma = \sqrt{H_2 H_6} \sigma$. Thus, at the enhan\c con
shell $r =r_{\rm e}$, these interior perturbations will satisfy the
boundary conditions
\begin{equation} \label{shellbc}
r_{\rm e} f'(r_{\rm e}) + (1 - \bar \sigma r_{\rm e} \coth \bar \sigma r_{\rm e}) f(r_{\rm e}) = 0.
\end{equation}

What we really want is boundary conditions for the exterior
perturbations $\Phi, \Psi_2, {\cal Z}_6$ at the shell. To obtain these
from the above conditions on $\Xi, \Gamma_2, {\cal H}_6$, we need to
work out the junction conditions at the shell for the perturbations,
and solve for $\Xi, \Gamma_2, {\cal H}_6$ and their first derivatives in
terms of $\Phi, \Psi_2, {\cal Z}_6$ and their derivatives.  

In general, the location of the shell separating the interior and and
exterior is also perturbed; however, we can use some of the remaining
diffeomorphism freedom in the ansatz to fix the coordinate position of
the shell to be $r_{\rm i} = r_{\rm e}$, the enhan\c con radius of the
unperturbed metric (by, say, an appropriate choice of the undetermined
constant $a$ in (\ref{rdif1},\ref{rdif2})). The dynamics of the shell
will then find its expression through the variation of the perturbed
metric at the shell location. This choice greatly simplifies the
problem of matching the perturbations at the shell.

Continuity of the metric and fields at the shell implies 
\begin{eqnarray}
\delta \phi(r_{\rm e}) = \delta \xi(r_{\rm e}), \quad \delta \psi_1(r_{\rm
e}) = \delta \gamma_1(r_{\rm e}), \quad \delta \psi_2(r_{\rm e}) = \delta
\gamma_2(r_{\rm e}), \\ \nonumber \delta Z_2(r_{\rm e}) = \delta H_2(r_{\rm e}), \quad 
\delta Z_6(r_{\rm e}) = \delta H_6(r_{\rm e}).
\end{eqnarray}
To relate the first derivatives, we compute the discontinuity in the
extrinsic curvature at the surface $r=r_{\rm e}$ when we patch the two
geometries together. This allows us to infer the stress tensor of the
perturbed shell from the supergravity point of view, with the result
\begin{equation}
S_{tt} = {1 \over 2\kappa^2 \sqrt{g_{rr}}} \left[ 2 {\bar Z_2' \over
\bar Z_2} - 2 {\bar Z_6' \over \bar Z_6} - 4 \psi_1' - \delta \psi_2'
- 2 {\bar H_2' \over \bar H_2} + 2 {\bar H_6' \over \bar H_6} + 4
\gamma_1' + \delta \gamma_2' \right] g_{tt},
\end{equation}
\begin{equation}
S_{\rho\sigma} = {1 \over 2\kappa^2 \sqrt{g_{rr}}} \left[ 2 {\bar Z_2'
\over \bar Z_2} - 2 {\bar Z_6' \over \bar Z_6} - 4 \psi_1' +{1 \over
2} \delta \psi_2' - 2 {\bar H_2' \over \bar H_2} + 2 {\bar H_6' \over
\bar H_6} + 4 \gamma_1' - {1 \over 2} \delta \gamma_2' \right] g_{\rho\sigma},
\end{equation}
\begin{equation}
S_{ij} = {1 \over 2\kappa^2 \sqrt{g_{rr}}} \left[  {\bar Z_2' \over
\bar Z_2} - 3 {\bar Z_6' \over \bar Z_6} - 4 \psi_1
-  {\bar H_2' \over \bar H_2} + 3 {\bar H_6' \over \bar H_6} + 4
\gamma_1' \right] g_{ij},
\end{equation}
\begin{equation}
S_{ab} = {1 \over 2\kappa^2 \sqrt{g_{rr}}} \left[  {\bar Z_2' \over
\bar Z_2} - 2 {\bar Z_6' \over \bar Z_6} - 4 \psi_1' 
-  {\bar H_2' \over \bar H_2} + 2 {\bar H_6' \over \bar H_6} + 4
\gamma_1' \right] g_{ab}
\end{equation}
(coordinates $\rho,\sigma$ run over $t,1,2$, $i,j$ over $\theta,\phi$,
and $a,b$ over the K3). We will assume that this shell stress tensor is
still sourced by a collection of BPS branes. Using the worldvolume
action for a wrapped D6-brane,
\begin{equation}
S = - \int_{{\cal M}_2} d^3 \xi e^{-\phi} (\mu_6 V(r) - \mu_2) (-\det
G_{\mu\nu})^{1/2} + \mu_6 \int_{{\cal M}_2 \times K3} C_{(7)} - \mu_2
\int_{{\cal M}_2} C_{(3)},
\end{equation}
where $G_{\mu\nu}$ is the induced string-frame metric, and the
string-frame volume $V(r) = V e^{\bar \phi-\psi_1} \bar Z_2 \bar Z_6^{-1}$,
one easily obtains the Einstein-frame stress-energy for a single %probe
brane in the exterior geometry,
\begin{equation}
S_{\mu\nu}^{brane} = - e^{3\bar \phi/4} (\mu_6 - \mu_2 V(r)^{-1}) g_{\mu\nu},
\end{equation}
\begin{equation}
S_{ab}^{brane} = - e^{3\bar \phi/4} \mu_6 g_{ab}
\end{equation}
(where $\mu,\nu$ run over $t,1,2$). If we use this to calculate the
stress-energy of the shell, we find that the value the stress tensor
should take can be written as
\begin{equation}
S_{\mu\nu}^{shell} = {1 \over 2\kappa^2 \sqrt{g_{rr}}}{e^{3\bar
\phi/4+\psi_1/4} \over \bar{Z}_2^{1/4} \bar{Z}_6^{-3/4} } \left[ {
Z_2' \over \bar Z_2} e^{-\bar \phi + \psi_1} + { Z_6' \over \bar Z_6}
\right] g_{\mu\nu},
\end{equation}
\begin{equation}
S_{ij}^{shell} = 0,
\end{equation}
\begin{equation}
S_{ab}^{shell} = {1 \over 2\kappa^2 \sqrt{g_{rr}}} {e^{3\bar \phi/4+\psi_1/4}
\over \bar{Z}_2^{1/4} \bar{Z}_6^{-3/4} } { Z_6' \over \bar Z_6} g_{ab}.
\end{equation}
The matching conditions for first derivatives of the metric are then
obtained by setting this brane stress tensor equal to the stress
tensor calculated from the supergravity point of view
above. Similarly, matching the discontinuity in the dilaton to the
brane source gives
\begin{equation}
\phi_{out}' - \phi_{in}' = {e^{3\bar \phi/4+\psi_1/4} \over 4 \bar
Z_2^{1/4} \bar Z_6^{-3/4}} \left[ e^{\psi_1 - \bar\phi} {Z_2' \over
\bar Z_2} - 3 {Z_6' \over \bar Z_6} \right].
\end{equation}
Taking the first-order part of all these equations gives us five
equations relating the derivatives of $\delta \phi, \delta \psi_1,
\delta \psi_2, \delta Z_2, \delta Z_6$ to the corresponding interior
quantities. 

We then have ten matching equations at the shell. However, we only
have nine quantities to specify: we want to solve for the three free
interior functions $\Xi, \Gamma_2, {\cal H}_6$ and their first
derivatives at the shell in terms of $\Phi, \Psi_2, {\cal Z}_6$ and
their first derivatives, and we also have three undetermined constants
in the diffeomorphisms to fix (since we already fixed one to satisfy
$r_i = r_{\rm e}$).\footnote{We have also checked explicitly that we
get the same boundary conditions at the shell for the physical degrees
of freedom even if we work in a coordinate system where the coordinate
location of the shell is not fixed.} Remarkably, this over-determined
system has a solution, and substituting into the boundary condition
(\ref{shellbc}) gives us, after considerable algebra, the relatively
simple expressions
\begin{eqnarray}
0&=&  -2 \hat r_6 (v^2+1) \Phi'(r_{\rm e}) - \hat r_6
 (v^2+4v-1)\Psi_2'(r_{\rm e}) + 4 \hat r_6 (v+1) {\cal Z}_6'(r_{\rm
 e}) \\ \nonumber &&+ v(v-1)^2 \Phi(r_{\rm e}) + (v-1)^2 {\cal Z}_6(r_{\rm e}),
\end{eqnarray}
\begin{eqnarray}
0 &=& -8 \hat r_6 (v+1) \Phi'(r_{\rm e}) +4 \hat r_6
 (v+1)\Psi_2'(r_{\rm e}) \\ && \nonumber + [4 \bar \sigma r_e \coth
 (\bar \sigma r_e) (v^2-1) -v^2 -6v+7] \Phi(r_{\rm e}) \\ &&+ [-2 \bar
 \sigma r_e \coth (\bar \sigma r_e) (v^2-1) +2 v^2 -2] \Psi_2(r_{\rm
 e})-3 (v-1)^2 {\cal Z}_6(r_{\rm e}), \nonumber
\end{eqnarray}
\begin{eqnarray}
0 &=& 4 \hat r_6 (v+1) \Phi'(r_{\rm e}) +4 \hat r_6 (v+1){\cal
 Z}_6'(r_{\rm e}) \\ && \nonumber + [-2 \bar \sigma r_e \coth (\bar
 \sigma r_e) (v^2-1) +v^2 +2v-3] \Phi(r_{\rm e}) \\&&+ [-2 \bar
 \sigma r_e \coth (\bar \sigma r_e) (v^2-1) +3 v^2 -2v-1] {\cal
 Z}_6(r_{\rm e}),\nonumber
\end{eqnarray}
where $v=V/V_*$ and $\bar \sigma = \sqrt{H_2 H_6} \sigma$. 

The problem of finding instabilities of the extremal enhan\c con
solution then reduces to looking for solutions of the equations of
motion for $\Phi, \Psi_2, {\cal Z}_6$ for some $\sigma$ which satisfy
these boundary conditions at the shell and fall off at large
distance. We have searched for such solutions using the same
relaxation techniques as in the previous section; we do not find any
instabilities. This result is, of course, what one would expect on the
basis of supersymmetry. A solution which satisfies the BPS bound may
have flat directions---it may be marginally stable to some
perturbation---but one would not expect that it will have any truly
unstable perturbations. 

\section{Conclusions}
\label{sec:concl}

In this paper, we have investigated the physics of non-extremal
enhan\c con solutions. We have found that the second branch of
non-extremal repulson solutions found in~\cite{JMPR:enh} appears to be
generically unphysical. The singularity in this solution cannot be
removed by excising the region inside the enhan\c con radius and
matching to a smooth interior across a physical enhan\c con shell. If
we attempt to impose such a matching, the shell required to achieve it
will violate the weak energy condition.

This result provides a resolution of a number of puzzling features of
these solutions observed in \cite{JMPR:enh}.  The first puzzle was
that the exterior solution never contains an event horizon, no matter
how large the non--extremality parameter $r_0$ became, in
contradiction to our expectation that the system would eventually
collapse to form a black hole. The second puzzle was that in the limit
of large K3 volumes, these solutions do not reproduce the expected
non-extremal $D6$--brane. In this limit the R--R $3$-form potential
vanishes as expected, but we still have additional dilaton hair. Since
the supergravity solutions with these features violate the weak energy
condition, these properties do not represent the real physics of
wrapped D6-branes.

In~\cite{JMPR:enh}, it was observed that this solution was presumably
just one member of a family of solutions with non-trivial dilaton
hair. We would expect that the problem we found here is quite generic
in such a family. One can interpret the violation of the weak energy
condition at the enhan\c con radius as saying that the energy in the
field configuration outside the enhan\c con exceeds the total ADM
energy. Thus, we need a negative-energy shell to make up the
deficit. We would therefore expect any solution with a lot of energy
in the dilaton outside the enhan\c con will similarly need a
negative-energy shell to avoid a a singularity, and would hence be
unphysical. Further investigations of more general non-extremal
supergravity solutions are in progress~\cite{dppr}.

We remarked earlier that this analysis also appears to extend to the
case with additional D2-brane charge. It would be interesting to
perform a similar analysis for the system of non-extremal fractional
branes considered in~\cite{obers}, where a similar two-branch
structure was found.

Investigating numerically the stability of the extremal enhan\c con,
we found that the shell is stable under small radial perturbations. We
used a perturbation ansatz, generalising the one adopted in
\cite{Maeda:stab}, which respected the symmetries of the unperturbed
solution. By fixing the diffeomorphism freedom, we were able to reduce
the linearized equations of motion to three coupled equations of
motion for the perturbed modes. We studied these equations numerically
for a wide range of the parameters $r_2$ and $r_6$, and failed to find
a solution which rendered the shell unstable. We concluded that at
least for radial perturbations, the extremal enhan\c con shell is
stable. This result is, of course, not a surprise: since this is a
supersymmetric solution, we would be very surprised to find an
instability. 

This result strengthens the argument of \cite{JMPR:enh} for the
consistency of the excision procedure \cite{JPP:enh}. This excision is
accomplished by the introduction of a shell of wrapped $D6$--branes at
the enhan\c con radius. The fact that this solution is stable is
another argument in favour of the excision and of the idea that the
solution is a sensible construction from the point of view of
supergravity.

The horizon branch of the non--extremal solution was also found to be
stable, in a numerical investigation for a wide range of the
parameters $\hat{r}_2$, $\hat{r}_6$ and $r_0$. Again we use the same
generalised ansatz and the diffeomorphism freedom to reduce to a
system of three coupled equations. We could not find any unstable
solutions to this configuration. This result is also fairly
unsurprising, since for large non--extremality the horizon branch
solution approaches a four dimensional Schwarzschild metric smeared
over a $K3$ and two longitudinal directions $x_1$, $x_2$. This
geometry is known to be stable against the kind of perturbations that
we are investigating.

What is quite interesting is that stability of the horizon branch
persists over the whole range of parameters. An instability of the
horizon branch could have been interpreted as evidence for the
existence of a new family of solutions, which might connect to the
extremal enhan\c con solution. While absence of evidence is not
evidence of absence, the stability of the horizon branch makes it seem
plausible that it provides the full description of non-extremal
enhan\c con physics for the range of parameters where it exists. 

What of small perturbations away from the extremal enhan\c con
solution, where there is no horizon branch? We do not have a
supergravity solution which describes them, but we do not feel this
implies some pathology in the physics. It may be that the appropriate
solutions lie outside the ansatz we have considered here; further
exploration of this possibility is in
progress~\cite{dppr}. Alternatively, it may be that the physics of
non-extreme enhan\c cons is not captured by a purely supergravity
solution. They could involve non-trivial non-abelian gauge fields, or
branes distributed in a `thick shell' over some finite range in the
radial coordinate. It will be very interesting to investigate this
question further. It would also be very interesting to investigate the
implications of these results for other applications of the enhan\c
con, such as \cite{JM,Constable,JJ:enh1}.

\medskip
\centerline{\bf Acknowledgements}
\medskip    
    
We are grateful to Rob Myers, Amanda Peet, Bernard Piette, Kostas
Kokkotas and Ioannis Floratos for useful discussions.  AD is supported in part
by EPSRC studentship 00800708 and by a studentship from the University
of Durham. SFR is supported by an EPSRC Advanced Fellowship.

\bibliographystyle{/home/dma3ad/Programs/TeX/utphys}  

\bibliography{apostolos}   

\providecommand{\href}[2]{#2}\begingroup\raggedright\begin{thebibliography}{10}

\bibitem{JPP:enh}
C.~V. Johnson, A.~W. Peet, and J.~Polchinski, ``Gauge theory and the excision
  of repulson singularities,'' {\em Phys. Rev. D} {\bf 61} (2000) 086001,
\href{http://www.arXiv.org/abs/hep-th/9911161}{{\tt hep-th/9911161}}.
%%CITATION = HEP-TH 9911161;%%.

\bibitem{buchel:enh}
A.~Buchel, A.~W. Peet, and J.~Polchinski, ``Gauge dual and noncommutative
  extension of an {$N=2$} supergravity solution,'' {\em Phys. Rev. D} {\bf 63}
  (2001) 044009,
\href{http://www.arXiv.org/abs/hep-th/0008076}{{\tt hep-th/0008076}}.
%%CITATION = HEP-TH 0008076;%%.

\bibitem{evans:enh}
N.~Evans, C.~V. Johnson, and M.~Petrini, ``The enhancon and {$N=2$} gauge
  theory/gravity {RG} flows,'' {\em JHEP} {\bf 10} (2000) 022,
\href{http://www.arXiv.org/abs/hep-th/0008081}{{\tt hep-th/0008081}}.
%%CITATION = HEP-TH 0008081;%%.

\bibitem{Behrndt:enh}
K.~Behrndt, ``About a class of exact string backgrounds,'' {\em Nucl. Phys.}
  {\bf B455} (1995) 188--210,
\href{http://www.arXiv.org/abs/hep-th/9506106}{{\tt hep-th/9506106}}.
%%CITATION = HEP-TH 9506106;%%.

\bibitem{Kallosh:enh}
R.~Kallosh and A.~D. Linde, ``Exact supersymmetric massive and massless white
  holes,'' {\em Phys. Rev. D} {\bf 52} (1995) 7137--7145,
\href{http://www.arXiv.org/abs/hep-th/9507022}{{\tt hep-th/9507022}}.
%%CITATION = HEP-TH 9507022;%%.

\bibitem{JMPR:enh}
C.~V. Johnson, R.~C. Myers, A.~W. Peet, and S.~F. Ross, ``The enhan\c con and
  the consistency of excision,'' {\em Phys. Rev. D} {\bf 64} (2001) 106001,
\href{http://www.arXiv.org/abs/hep-th/0105077}{{\tt hep-th/0105077}}.
%%CITATION = HEP-TH 0105077;%%.

\bibitem{es1}
A.~Dimitriadis and S.~F. Ross, ``Stability of the non-extremal enhancon
  solution. {I}: Perturbation equations,'' {\em Phys. Rev. D} {\bf 66} (2002)
  106003,
\href{http://www.arXiv.org/abs/hep-th/0207183}{{\tt hep-th/0207183}}.
%%CITATION = HEP-TH 0207183;%%.

\bibitem{wijnholt}
M.~Wijnholt and S.~Zhukov, ``Inside an enhancon: Monopoles and dual
  {Yang-Mills} theory,'' {\em Nucl. Phys.} {\bf B639} (2002) 343--369,
\href{http://www.arXiv.org/abs/hep-th/0110109}{{\tt hep-th/0110109}}.
%%CITATION = HEP-TH 0110109;%%.

\bibitem{Maeda:stab}
K.~Maeda, T.~Torii, M.~Narita, and S.~Yahikozawa, ``The stability of the shell
  of {D2-D6} branes in a {$N = 2$} supergravity solution,'' {\em Phys. Rev. D}
  {\bf 65} (2002) 024030,
\href{http://www.arXiv.org/abs/hep-th/0107060}{{\tt hep-th/0107060}}.
%%CITATION = HEP-TH 0107060;%%.

\bibitem{PPT:num}
W.~H. Press, S.~A. Teukolsky, W.~T. Vetterling, and B.~P. Flannery, {\em
  Numerical Recipes in {C}$++$: The Art of Scientific Computing}.
\newblock Cambridge Univ. Press, Cambridge, UK, 2002.

\bibitem{chandra}
S.~Chandrasekhar, {\em The Mathematical theory of black holes}.
\newblock Oxford Univ. Press, New York, Oxford, 1992.

\bibitem{GL}
R.~Gregory and R.~Laflamme, ``Black strings and p-branes are unstable,'' {\em
  Phys. Rev. Lett.} {\bf 70} (1993) 2837--2840,
\href{http://www.arXiv.org/abs/hep-th/9301052}{{\tt hep-th/9301052}}.
%%CITATION = HEP-TH 9301052;%%.

\bibitem{dppr}
A.~Dimitriadis, A.~W. Peet, G.~Potvin, and S.~F. Ross. Paper in preparation.

\bibitem{obers}
M.~Bertolini, T.~Harmark, N.~A. Obers, and A.~Westerberg, ``Non-extremal
  fractional branes,'' {\em Nucl. Phys.} {\bf B632} (2002) 257--282,
\href{http://www.arXiv.org/abs/hep-th/0203064}{{\tt hep-th/0203064}}.
%%CITATION = HEP-TH 0203064;%%.

\bibitem{JM}
C.~V. Johnson and R.~C. Myers, ``The enhancon, black holes, and the second
  law,'' {\em Phys. Rev. D} {\bf 64} (2001) 106002,
\href{http://www.arXiv.org/abs/hep-th/0105159}{{\tt hep-th/0105159}}.
%%CITATION = HEP-TH 0105159;%%.

\bibitem{Constable}
N.~R. Constable, ``The entropy of {4D} black holes and the enhancon,'' {\em
  Phys. Rev. D} {\bf 64} (2001) 104004,
\href{http://www.arXiv.org/abs/hep-th/0106038}{{\tt hep-th/0106038}}.
%%CITATION = HEP-TH 0106038;%%.

\bibitem{JJ:enh1}
L.~Jarv and C.~V. Johnson, ``Rotating black holes, closed time-like curves,
  thermodynamics, and the enhancon mechanism,'' {\em Phys. Rev. D} {\bf 67}
  (2003) 066003,
\href{http://www.arXiv.org/abs/hep-th/0211097}{{\tt hep-th/0211097}}.
%%CITATION = HEP-TH 0211097;%%.

\end{thebibliography}\endgroup
    
\end{document}